\documentclass[aps,prb,showpacs,twocolumn,superscriptaddress]{revtex4-1}

\usepackage{amsmath}
\usepackage{amsfonts}
\usepackage{braket}
\usepackage{amssymb}
\usepackage{graphicx}
\usepackage{natbib}
\usepackage{bbold}
\usepackage{bm}
\usepackage[colorlinks=true,linkcolor=blue,urlcolor=black,citecolor=blue]{hyperref}

\DeclareGraphicsExtensions{.pdf,.png}

\newcommand{\ti}{\tilde}

\begin{document}
\title{Spatial compression of a particle state in a parabolic potential by spin measurements. }

\author{P. V. Pyshkin}
\affiliation{Beijing Computational Science Research Center, Beijing 100094, China}
\affiliation{Department of Theoretical Physics and History of Science, University of the Basque Country UPV/EHU, 48080 Bilbao, Spain}
\affiliation{Ikerbasque, Basque Foundation for Science, 48011 Bilbao, Spain}

\author{E. Ya. Sherman}
\affiliation{Department of Physical Chemistry, University of the Basque Country UPV/EHU, 48080 Bilbao, Spain}
\affiliation{Ikerbasque, Basque Foundation for Science, 48011 Bilbao, Spain}

\author{Da-Wei Luo}
\affiliation{Beijing Computational Science Research Center, Beijing 100094, China}
\affiliation{Department of Theoretical Physics and History of Science, University of the Basque Country UPV/EHU, 48080 Bilbao, Spain}
\affiliation{Ikerbasque, Basque Foundation for Science, 48011 Bilbao, Spain}

\author{J. Q. You}
\affiliation{Beijing Computational Science Research Center, Beijing 100094, China}

\author{Lian-Ao Wu}
\email{lianao.wu@ehu.es}
\affiliation{Department of Theoretical Physics and History of Science, University of the Basque Country UPV/EHU, 48080 Bilbao, Spain}
\affiliation{Ikerbasque, Basque Foundation for Science, 48011 Bilbao, Spain}

\date{\today}

\begin{abstract}
	We propose a scheme for engineering compressed spatial states in a two-dimensional parabolic potential 
	with a spin-orbit coupling by selective spin measurements. This sequence of measurements results in a 
	coordinate-dependent density matrix with probability maxima achieved at a set of lines or at a two dimensional lattice.
	The resultant probability density depends on the spin-orbit coupling and the potential parameters
	and allows one to obtain a broad class of localized pure states on demand. The proposed scheme can be realized in spin-orbit 
	coupled Bose-Einstein condensates. 
\end{abstract}


\maketitle

\section{Introduction} 

In recent years, quantum control of individual particles and ensembles has attracted a lot of attention. 
Experimentally, quantum dots (QD)~\cite{rev_qd}, ion traps~\cite{trapped_review_1}, and Bose-Einstein 
condensates(BEC)~\cite{bec_book,bec-qc} are promising candidates for designing new nano- and micro-elements 
for quantum technologies~\cite{Nielsen2000}. Of particular interest is the possibility of 
simultaneous control and manipulation of degrees of freedom of different origin such as position and spin by means 
of spin-orbit coupling present in broad variety of solid-state and condensed matter systems ~\cite{rashba_2016}. For instance,
spin and position of electron in semiconductor-based QDs may be mutually related due to this coupling. 
A synthetic SOC realized in BEC~\cite{bec_so_1} demonstrate a variety of new phenomena, not achievable in solid state-based system.   
One of advantages of SOC is that it is externally tunable,  either by a static electric field for semiconductors  \cite{soc_book}
or by optical fields in the condensates~\cite{bec_so_1}.

In many applications, it is important to control a quantum system in a binding potential and prepare on demand a 
quantum state, for example, initial pure states of qubits for quantum computation. Various methods of 
quantum system preparations~\cite{note1} such as logical-~\cite{Chuang_logical_labeling} or temporal 
labeling~\cite{Chuang_temporal_labeling} were proposed. Production of coherent state of an ion in a parabolic potential has
also been shown to be feasible in an ion trap~\cite{ion_coherent}. Due to the high interest in circuit quantum 
electrodynamics and ion trap experiments, spatially localized states are important for the particle manipulation (see for example~\cite{newadd_Lo, newadd_Owen, newadd_Douglas}). 
A widely used approach to generate and control the quantum state of a target system is by designing 
a set of tailored {\em selective} (where the outcome is either discarded or accepted, 
dependent on the measurement result) 
measurements of an ancilla coupled to the target. 
This approach has been theoretically proposed for pure state generation and ground state 
cooling~\cite{Hiromichi_Main_method, Li2011}, entanglement generation~\cite{Wu04}, 
and experimentally realized in~\cite{Xu2014} for photons. In this paper, we propose a 
tailored spin measurement to spatially compress the quantum state of a particle with SOC 
in a two-dimensional (2D) parabolic potential. As examples, two types of compression schemes 
are studied, where the state is compressed along a certain direction (\textit{linear compressed state}) 
or into a two-dimensional lattice similar to \textit{Lotus-seed} probability distribution. Experimental 
tolerance against measurement time errors is also analyzed.

The paper is organized as follows. In Section II we show how 
selective spin measurements on a 2D system with SOC can produce a spatially compressed stripes-like density
distribution. In Section III we consider one-dimensional 
(1D) realization of this compression. In Section IV we show that spatial compression can result
in a ``probability lattice'' which we call a ``Lotus-seed state''. In Sections V and VI we consider efficiency 
of the proposed technique and discuss the results. Section VII presents the conclusions. 
Appendix contains some mathematical details and extra figures.

\section{Spatial compression by spin measurements} We begin with considering a pseudo spin-$1/2$ particle in a 2D parabolic potential described 
by the Hamiltonian \cite{Hamiltonian_proof_1,Hamiltonian_proof_2}
\begin{equation}\label{main_H}
H = \frac{p_x^2 + p_y^2}{2M} + \frac{M\omega^2}{2}(x^2+y^2) + \alpha (\bm{p\cdot n})(\bm{\sigma\cdot m}),
\end{equation}
where $\bm{p}$ is the momentum operator, $\bm{n}$ and $\bm{m}$ are unit vectors ($\bm{n}$ lies in the $XY$ plane), 
$\bm{\sigma}=(\sigma_x, \sigma_y, \sigma_z)$ is the Pauli matrices vector, $M$ is the particle mass, $\omega$ 
is the oscillator frequency, and $\alpha$ is the SOC constant. The Hamiltonian can be diagonalized via a 
unitary transformation $\mathcal U(\alpha)= \cos\chi - i(\bm{\sigma\cdot m})\sin\chi$,
where $\chi\equiv\alpha M\hbar^{-1} (\bm{r\cdot n}) $. The diagonalized Hamiltonian $H' = \mathcal U^\dagger(\alpha) H \mathcal U(\alpha)$ 
reads
\begin{equation}\label{H_rotated}
H'=\left(\hbar\omega(a^\dagger a+b^\dagger b +1) - \frac{\alpha^2 M}{2}\right)\otimes\mathbf{1} ,
\end{equation}
where $a^\dagger(b^\dagger)$ and $a(b)$ are the raising and lowering operators for $x$($y$) directions. 
It's clear that the presence of SOC causes an energy shift $-\alpha^2 M/2$.

Now we propose a method to spatially compress a particle state by implementing selective measurements 
on the spin degree of freedom~\cite{note_spin_measurement}. In general, the spin measurement can be 
described by a rotated basis $\{\ket{+}, \ket{-}\}$, where $\ket{+}=\cos\theta/2\ket{\uparrow}+e^{i\varphi}\sin\theta/2\ket{\downarrow}$, 
$\ket{-}=\sin\theta/2\ket{\uparrow}-e^{i\varphi}\cos\theta/2\ket{\downarrow}$, with $\theta$ and $\varphi$ 
being the polar and azimuthal angles, and $\ket{\uparrow}, \ket{\downarrow}$ denote the spin up and down states. 
At $N$ evenly-spaced times $t=t_1, t_2, \dots, t_N$, we make {\em selective} measurements described by projection operator 
$\mathbb{1}\otimes\ket{+}\bra{+}$ on the spin state of a particle, which plays the role of an ancilla. 
After each measurement, we discard the~$|- \rangle$ result, thus producing a selective measurement. 
Between every two consecutive spin measurements, the particle 
evolves according to the Hamiltonian~(\ref{main_H}). For an initial particle state described by a 
factorized density matrix $\rho_0=\rho_{c0}\otimes\ket{+}\bra{+}$, where $\rho_{c0}$ describes 
the position dependence, after~$N$ selective spin measurements, the final density matrix becomes
\begin{eqnarray}
&&\rho_N = \rho_{cN}\otimes\ket{+}\bra{+}, \;\; \rho_{cN} = \frac{V^N\rho_{c0}V^{\dagger N}}{P(N)}, \label{rho_transform_1} \nonumber\\
&& V\equiv\bra{+}U(\tau)\ket{+},\label{V_definition}
\end{eqnarray}
where $P(N)\equiv \mathrm{Tr}(V(\tau)^N\rho_{c0}V^{\dagger N}(\tau))$ is the corresponding survival 
probability of the system, $U(\tau) = \exp(-iH\tau/\hbar)$ is the evolution operator, 
$\tau = t_{i+1}-t_i$ is the interval between two consecutive measurements, assuming to be instantaneous. We show that by selecting $\tau$ in a certain manner, 
one can obtain the final~$\rho_N$ corresponding to {\em {spatially localized state}} 
of a special kind in the limit~$N\rightarrow\infty$.


After some algebra~(see Appendix) we obtain
\begin{align}\label{V_2}
V(\tau) =& \left[ \vphantom{\frac{1}{1}} (\cos\chi-if\sin\chi)(\cos\chi'+if\sin\chi')  \right. \nonumber\\
&+\left. |g|^2\sin\chi\sin\chi' \vphantom{\frac{1}{1}} \right] e^{-i\ti{H}\tau/\hbar},
\end{align}
where $\chi' = e^{-i\ti{H}\tau/\hbar}\chi e^{i\ti{H}\tau/\hbar}$, $\ti{H}$ is the spatial part 
of $H'$, $f\equiv\braket{+|(\bm{\sigma\cdot m)}|+}$ and $g\equiv\braket{+|(\bm{\sigma\cdot m)}|-}$. 

It is noticeable that while $V$ is hard to be calculated even numerically~\cite{Wu04}, 
we can have analytical solutions in the following two cases:\\
1) $\omega\tau = 2\pi j$, when $\chi' = \chi$; \\
2) $\omega\tau = \pi + 2\pi j$, when $\chi' = - \chi$,\\
where $j's$ (= 0,1,2,\dots) are integers. It is easy to see that in the first case we 
have $V(\tau)=e^{-2\pi i j\ti{H} /\hbar\omega}$, which is not of interest because $V(\tau)$ 
acts on the oscillator states as unity operator. Henceforth we will consider the second case.

To be concrete we consider a realization with $\theta=0$ and $\bm{m}=(m_x,m_y,0)$ and obtain
\begin{equation}\label{V_3}
V(\tau) = \cos(2\chi)e^{-2\pi i\ti{H}(j+1/2)/\hbar\omega}.
\end{equation}
The eigenstates of operator~$e^{-2\pi i\ti{H}(j+1/2)/\hbar\omega}$ are number states with 
eigenvalues ~$A(-1)^k$(see \cite{eigen}), where $A=\exp(2\pi i(j+1/2)[\alpha^2M/2\hbar\omega - 1])$ and $k=k_x+k_y$ is 
the total number of excitations related to the~$x$ and~$y$ spatial degrees of freedom. 
Therefore, when we substitute Eq.~(\ref{V_3}) in Eq.~(\ref{V_definition}), we 
can keep only~$\cos(2\chi)$ in~$V(\tau)$ when (i) the initial state $\rho_{c0}$ is a 
pure number state $\rho_{c0}=\ket{k_x,k_y}\bra{k_x,k_y}$, or (ii) if the initial 
mixed state is diagonal in the number states basis: $\rho_{c0} = \sum_{N=0}^\infty\sum_{k_x=0}^N p(N, k_x)\ket{k_x,N-k_x}\bra{k_x,N-k_x}$,  
exemplified by the thermal state, where $p(N,k_x)$ are classical probabilities and ket (bra) 
vector contains number of excitations related to the $x$ and~$y$ degrees of freedom. 
In the coordinate representation, our result becomes
\begin{align}\label{rho_transform_coo}
\rho&_{cN}(x,y;x',y') = \nonumber\\ 
&\cos^N(2\chi(x,y))\rho_{c0}(x,y;x',y')\cos^N(2\chi(x',y')) \nonumber \\
&\times \left(\int_{-\infty}^{+\infty}\int_{-\infty}^{+\infty}\cos^{2N}(2\chi(x,y))\rho_{c0}(x,y;x,y)dxdy     \right)^{-1}.
\end{align}
As can be seen from Eq.~(\ref{rho_transform_coo}), the resultant density matrix~$\rho_{cN}(x,y;x',y')$ 
has a sharp maxima in regions where $|\cos(2\chi(x,y))|=1$. These regions correspond to lines in the~$xy$-plane:
\begin{equation}\label{line_l}
(\bm{r\cdot n}) = \frac{\pi l \hbar}{2\alpha M}, \; l=0,\pm1,\pm2, \dots,
\end{equation}
allowing us to achieve the desired {\em lower-symmetry spatial compression}. Particularly if the 
SOC is weak (i.e., $\alpha\ll\hbar d^{-1}M^{-1}$, where $d$ is the size of the system), 
then it is sufficient to take only~$l=0$ in Eq.~(\ref{line_l}). Distance between lines (\ref{line_l}) corresponds to spin-precession length~$\hbar/M\alpha$~(see Ref.\cite{Winkler}) and does not depend
on~$\omega$.

\begin{figure}[t]
\begin{center}
\includegraphics{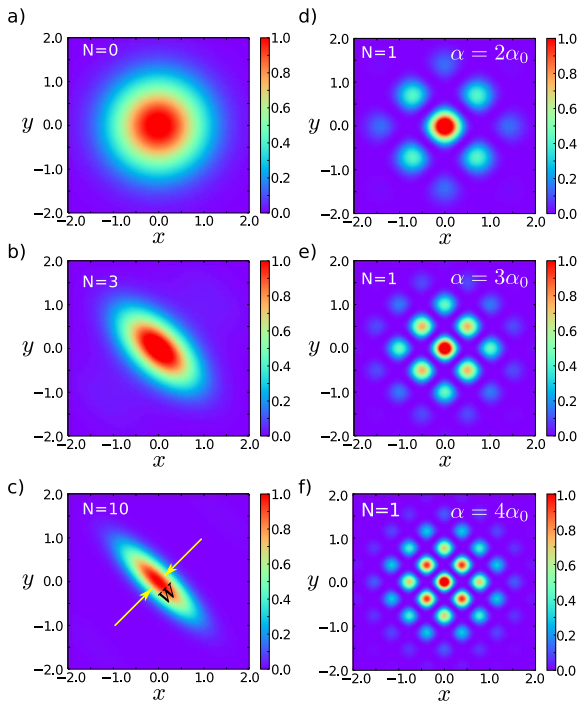}
\end{center}
\caption{Density of the spatial probability of finding particle after several 
successful measurements. Left column: ``linear compressed state''.  
The probability density of initial ground state (a), after $N=3$ (b) and $N=10$ (c) 
successful measurements. In~(c) we additionally illustrate definition of half-width~$W$~(see Eq.~\eqref{halfwidth}). 
Right column: ``Lotus-seed state''. We display the probability distributions after two successful 
spin measurements -- one before changing electric field and another one after, 
with $\alpha = 2\alpha_0$ (d), $\alpha = 3\alpha_0$ (e) and $\alpha = 4\alpha_0$ (f).}
\label{fig1}
\end{figure}

Unlike~\cite{Hiromichi_Main_method, Li2011}, the resultant state, Eq.~(\ref{rho_transform_coo}), 
of our approach does not increase purity of the initial spatially-symmetric (thermal) state of 
the particle to~$1$, even if~$N\rightarrow\infty$. 
However, it can be used for obtaining pure states out of {\it non-symmetric} mixed states. For example, 
if spatial part of the initial state is given by density matrix $\rho_{c0} = \ket{0,0}\bra{0,0}/2 + \ket{0,1}\bra{0,1}/2$, 
with the purity $\mathrm{Tr}\rho_{c0}^2 = 1/2$, then for $\bm{n}=(0,1,0)$, we obtain $\mathrm{Tr}\rho_{cN}^2 \rightarrow 1$ 
at $N\rightarrow\infty$~\cite{purity}.

To illustrate our proposal, we consider a realization~$\alpha = \alpha_0 = \sqrt{\hbar\omega/2M}$ and~$\bm{n}=(1/\sqrt2, 1/\sqrt2, 0)$. 
Initial state for the particle is the ground state~$\rho_{c0} = \ket{0,0}\bra{0,0}$. 
In Figs.~\ref{fig1}(a)-\ref{fig1}(c) we present spatial probability densities~$\rho_{cN}(x,x;y,y)$, 
where progressively more compressed state is observed for higher number of successful measurements 
(see \cite{Wu_spin_measurement} and references therein, e.g.,  experimental implementation of spin measurements). 

This setting can be realized in a QD (with a typical $\hbar\omega\approx1$~meV) located in an InSb 
semiconductor layer~\cite{Winkler} with a width of about~$10^{-6}$~cm in a perpendicular electric 
field of ~$~10^5$~V/cm.  In this case, the general Hamiltonian with both 
Dresselhaus~\cite{Dresselhaus} and Rashba~\cite{Rashba} terms can be expressed as~(\ref{main_H}). 
However, this QD-based realization of the compression is difficult in terms of the frequency and timing accuracy, which is 
required to be higher than 1~ps.  
Therefore, atomic systems may be much better suitable for obtaining the proposed spatially compressed states. 
For example, the characteristic frequency $\omega$ of trapped BEC is only 
about $300$~s$^{-1}$ and achievable {\em synthetic} spin-orbit coupling strength 
is $\alpha\approx4\alpha_0$ (Ref. [\onlinecite{bec_so_1}]). Projective pseudo-spin measurements on 
BEC have been used recently in Ref. [\onlinecite{bec_spin_measurement}].

\section{Compression in one-dimensional systems.} 

\begin{figure}[t]
	\begin{center}
		\includegraphics{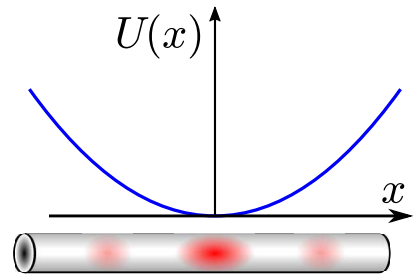}
	\end{center}
	\caption{Schematic illustration of 1D quantum wire with parabolic potential~$U(x)$. Spatial electron density produced by the measurements is illustrated by red color inside the	wire. }
	\label{fig_wire}
\end{figure}

Our proposal can also be applied to a one-dimensional system (such as a quantum wire or a tightly compressed BEC) 
in a parabolic potential with spin-orbit interaction~\cite{SOC_in_wire}, schematically shown in~Fig.~\ref{fig_wire}. 
In this case, the Hamiltonian~(\ref{main_H}) can be rewritten as 
\begin{equation*}
H = \frac{p^2}{2M} + M\omega^2 \frac{x^2}{2} + \alpha p (\bm{\sigma\cdot m}).
\end{equation*}
Note that in the 1D case, this Hamiltonian is valid for {\it any ratio} of Rashba and Dresselhaus coupling strengths 
which determine the direction of vector~$\bm{m}$. All results for the wire are similar to those for~2D systems 
(we again choose $\bm{m} = (m_x, m_y, 0)$, $\theta = 0$), and Eq.~(6) can be rewritten as 
\begin{align*}
\rho&_{cN}(x;x') = \cos^N(2\chi(x))\rho_{c0}(x;x')\cos^N(2\chi(x')) \nonumber \\
&\times \left(\int_{-\infty}^{+\infty}\cos^{2N}(2\chi(x))\rho_{c0}(x;x)dx     \right)^{-1},
\end{align*}
where~$\chi\equiv\alpha M x/\hbar.$ Instead of Eq.~(\ref{line_l}) for the 2D potential, 
in~1D we obtain the density distributed near special {\em points} along the wire: $x_l = \pi l \hbar/2\alpha M,\; l=0,\pm1, \pm2, \dots$.

\section{Compression into a ``Lotus-seed state''} Another interesting possibility is given by 
changing the SOC term in Hamiltonian~(\ref{main_H}) after $N$ successful spin measurements and 
performing $N$ further selective spin measurements with a different SOC term. Physically, 
it can be realized, for example, by changing the applied electric field across the QD plane to the opposite 
direction, i.e., after first $N$ measurements we change the vector $\bm{n}$ in Hamiltonian~(\ref{main_H}) 
to $\bm{n'}$, where $\bm{n'}\perp\bm{n}$. In this manner the final spatial probability distribution will 
be different. Since spatial compression with a given~$\bm{n}$ generates probability lines in Eq.~(\ref{line_l}), 
it is easy to understand that by changing~$\bm{n}$ we obtain a state with {\em Lotus-seed spatial distribution} 
(Figs.~\ref{fig1}(d)-\ref{fig1}(f)). Here, peaks in the probability density correspond to the intersections of 
probability lines in Eq.~(\ref{line_l}) for different~$\bm{n}$ and~$\bm{n'}$. In Figs.~\ref{fig1}(d)-\ref{fig1}(f) 
we show the density of the spatial probability for the same initial ground state as before, $\bm{m}$ and~$\bm{n}$ 
as in the previous example for~$\alpha = 2\alpha_0$ [Fig.~\ref{fig1}~(d)], $\alpha = 3\alpha_0$ [Fig.~\ref{fig1}~(e)] 
and~$\alpha = 4\alpha_0$ [Fig.~\ref{fig1}~(f)]. Here, after the first ($N=1$) selective measurement we change the 
direction of~$\bm{n}$ and after another successful spin measurement with~$\bm{n'}$ we obtain a  
Lotus-seed spatial distribution of electron spatial density. Note that states shown in Figs.~\ref{fig1}(d)-\ref{fig1}(f) are pure.

\begin{figure}
\begin{center}
\includegraphics{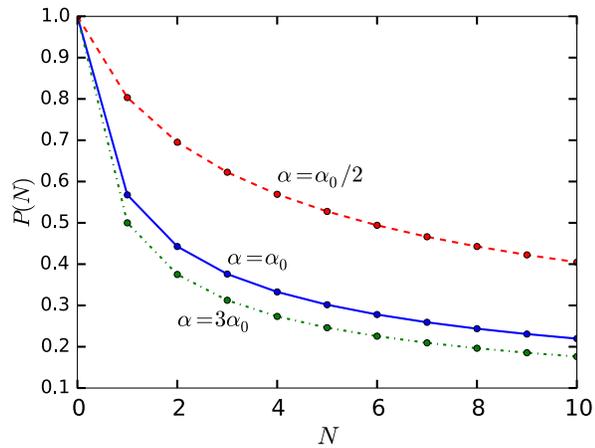}
\end{center}
\caption{Survival probability for ``linear compressed state'' case: $\alpha = \alpha_0/2$ (dashed red line), 
$\alpha = \alpha_0$ (solid blue line) and $\alpha = 3\alpha_0$ (dotted green line).}
\label{fig2}
\end{figure}

\section{Efficiency of spatial compression} In order to evaluate the effectiveness of the 
spatial compression we can calculate characteristic size of the region with essential 
probability density as a function of the number of successful measurements~$N$. 
In the case of ``linear compression'' (Eq.~(\ref{rho_transform_coo}) and Fig.~\ref{fig1}) 
and for large~$N\gg\hbar\omega/M\alpha^2$ we obtain half-width~$W(N)$ [see Fig.~\ref{fig1}~(c)] 
of the spatial probability function as~(see Appendix):
\begin{equation}\label{halfwidth}
W(N) = \frac{\hbar}{\alpha M \sqrt{2N}}\propto\frac{1}{\sqrt{N}}.
\end{equation}
As expected, $W(N)$ tends to zero as $N\rightarrow\infty$. In Fig.~\ref{fig2}, we present 
the dependence of survival probability~$P(N)$ defined below Eq.~(\ref{V_definition}) on the number of measurements for three 
different values of the SOC constant for ``linear compression'' (``Lotus-seed state'' plot is presented in the Appendix).

\begin{figure}
\begin{center}
\includegraphics{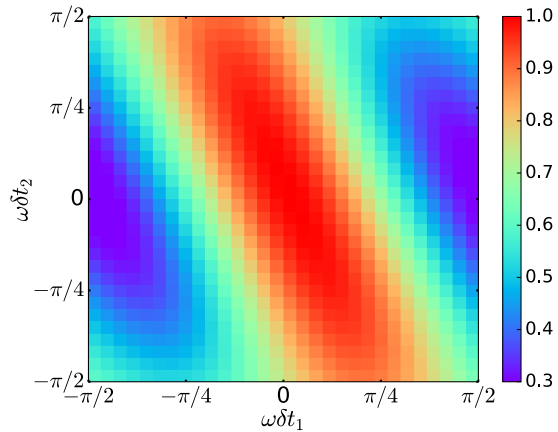}
\end{center}
\caption{Fidelity between the ideal Lotus seed state ($\alpha=\alpha_0$,  two spin measurements 
after two evolutions with different vectors $\bm{n}_1, \, \bm{n}_2$, $\bm{n}_1 \perp \bm{n}_2 $) 
and another spatial quantum state resulted from two spin measurements with time 
shifts~$\delta t_1$ and $\delta t_2$ from the ideal time $\pi/\omega$.}
\label{fig3}
\end{figure}

Experimentally, it is important to ensure the robustness against the measurement errors. 
As we have noticed, Eq.~(\ref{rho_transform_coo}) is valid only when all~$N$ measurements are 
performed with ideal intervals $\tau = 2\pi\omega^{-1}(1/2 + j)$ between each of two consecutive measurements. 
To achieve a high fidelity, the time interval~$\tau$ can be different from~$\pi/\omega$, but the 
absolute time error~$\delta t$ must be small: $\delta t \ll \pi/\omega$. In Fig.~\ref{fig3} 
we present the time dependence of fidelity~$F$~\cite{Fidelity} between ideal lotus seed state achieved by 
sequentially applying Eq.~(\ref{rho_transform_coo}) with different~$\chi$ and spatial state which results 
from two measurements made at non-ideal times~$\tau_{1,2} = \pi/\omega + \delta t_{1,2}$.
The central point of Fig.~\ref{fig3} corresponds to two sequential ideal spin measurements 
where~$F(0,0)=1$. As can be seen from~Fig.~\ref{fig3}, $F$ is tolerant against small time errors. 
Asymmetry with respect to $\delta t_1$ and $\delta t_2$ in Fig.~\ref{fig3} is due to the dependence 
of  time errors contribution to the final state on the step number.

\section{Discussion} 

As mentioned above, the resultant spatial wave function can be expressed as a superposition 
of high-energy oscillator states (each non-discarded spin measurement increases the energy 
by the value of the order of~$M\alpha^2$). However, the ideal oscillator spectrum may 
not be realizable in realistic systems. A QD has only a few low-lying levels with almost equal 
energy gaps. Thus, the electron spatial probability will certainly be less sharply defined 
than in Fig.~\ref{fig1}. This also means that there is no sense in performing many measurements 
if the SOC constant is of the order~$\alpha_0$ or larger.

Note that if the last term in Eq.~(\ref{main_H}) is replaced by~$\alpha' (\bm{r\cdot n})(\bm{\sigma \cdot m})$ for a
a spin-coordinate coupling, we can repeat all our calculations and achieve the same results by swapping variables $\bm{p}$ and $\bm{r}$. 
In this case we can generate the compressed states in the momentum space instead of spatial compression described above. 
This can be realized, for example, by means of an inhomogeneous Zeeman field.

In addition to the spatial compression, by using only single-shot selective spin measurement we can 
achieve a Schr\"odinger cat like spatial superposition state~\cite{cat_state}(see Appendix). 
If the initial state is the ground state with spin~$\ket{\uparrow}$ then, as follows from Eq.~(\ref{main_H}), 
this state is a superposition of two states (eigenstates of $(\bm{\sigma\cdot m})$) which have opposite 
velocities $\pm\alpha/\sqrt2$ with directions collinear to~$\bm{n}$~\cite{Sherman}. Therefore, by
performing spin measurement at time~$\pi/4\omega$,  we can achieve separated peaks in the  
spatial probability distribution ~(see Appendix). The distance~$L$ between these two peaks 
can be estimated as $L\approx(\alpha/\alpha_0)l_0$, where $l_0=\sqrt{\hbar/M\omega}$ is the oscillator length. This distance corresponds to the spin
separation length~\cite{Sherman} arising due to anomalous spin-dependent
velocity ($\propto\alpha$) on the timescale of the measurement and can
be approximately written as $L \approx \alpha (\pi/2\omega)$.

For completeness, we mention that our proposal is robust against decoherence effects
when the condition~$N\tau<T_d$, where $T_d$ is the 
characteristic decoherence time, holds. This condition, which can be reformulated as $\omega\,T_{d}\gg1,$ meaning that the 
orbital states are well-defined, holds for the majority of localized quantum systems weakly coupled to the 
environment \cite{about_relax_time,about_relax_time_trap}.


\section{Conclusion} We proposed a technique for achieving two kinds of spatially 
compressed states in two-dimensional harmonic potential with spin-orbit coupling. It has been shown that 
tailored {\em selective} projective measurements of the spin of the particle can dramatically modify the position-dependent probability 
density. As a result, one can create \textit{on demand} a variety of stripe- and Lotus-seed-like density distributions, 
dependent on the measurement protocols and system parameters. We suggest that spin-orbit coupled Bose-Einstein 
condensates are suitable for realization of this proposal.

{\acknowledgments We acknowledge grant support from the Basque Government (grant IT472-10), the Spanish MINECO/FEDER 
(No. FIS2012-36673-C03-03, No. FIS2012-36673-C03-01), the NBRPC No.~2014CB921401, the NSAF No. U1330201, the NSFC No. 91421102, and University of Basque Country
UPV/EHU under program UFI 11/55. We acknowledge Prof. I. Spielman for consultation on Ref. [\onlinecite{bec_so_1}.] 

\section*{Appendix}
\appendix*

\subsection{Derivation of Equation (4) using the ladder operators.}

Using the diagonalized Hamiltonian in Eq.~(2), we have
\begin{equation}\label{V_1}
V(\tau)=Fe^{-i\ti{H}\tau/\hbar}F^\dag +  Ge^{-i\ti{H}\tau/\hbar}G^\dag.
\end{equation}
Here $\ti{H}$ is a spatial part of $H'$ which does not act on the spin degree of freedom, 
$F \equiv \cos\chi - if\sin\chi$, $G \equiv - ig\sin\chi$, 
and $f\equiv\braket{+|(\bm{\sigma\cdot m)}|+}$, $g\equiv\braket{+|(\bm{\sigma\cdot m)}|-}$.

For obtaining expression (4) in the main text, we use the relation
\begin{equation}
V(\tau) = V(\tau)e^{i\ti{H}\tau/\hbar}e^{-i\ti{H}\tau/\hbar},
\end{equation}
and the following property of a unitary operator:
\begin{equation}
e^{-i\ti{H}\tau/\hbar}f(\hat{\chi})e^{i\ti{H}\tau/\hbar} = f(e^{-i\ti{H}\tau/\hbar}\hat{\chi}e^{i\ti{H}\tau/\hbar}),
\end{equation}
where the operator $\hat\chi$ depends on the ladder operators $a^\dagger, a, b^\dagger, b$, 
and the operator function $f$ in our case is $\sin\hat\chi$ or $\cos\hat\chi$. 

With the above transformations, we can define a new operator
\begin{equation}\label{chi_prime}
\chi' = e^{-i\ti{H}\tau/\hbar}\chi e^{i\ti{H}\tau/\hbar}.
\end{equation}
The operator $\chi$ is a function of coordinate operators $a^\dagger + a$ and $b^\dagger + b$.

To simplify expression for $\chi'$, we use the known property of ladder operators:
\begin{equation} 
e^{\gamma a^\dagger a} a  e^{-\gamma a^\dagger a} = e^{-\gamma} a, \;\;\;  e^{\gamma a^\dagger a} a^\dagger  e^{-\gamma a^\dagger a} = e^{\gamma} a^\dagger.
\end{equation}
If $\gamma$ in the above relations is $2\pi i j$ ($j = 0,1,2,\dots $), then it means that:
\begin{equation}
e^{2\pi i j a^\dagger a} (a^\dagger + a)  e^{-2\pi i j a^\dagger a} = a^\dagger + a.
\end{equation}
The above equation corresponds to the case $\chi = \chi'$ in the main text. If now $\gamma$ in the above relations is $2\pi j + \pi$, then
\begin{equation}
e^{2\pi i (j+1/2) a^\dagger a} (a^\dagger + a)  e^{-2\pi i (j+1/2) a^\dagger a} = - (a^\dagger + a),
\end{equation}
corresponding to the $\chi = - \chi'$ case which is of interest to us. Note that only for these 
two cases the unitary transformation~(\ref{chi_prime}) transforms one coordinate to another, while for other values of $\gamma$, coordinates are transformed into a function of coordinates and momenta. 

\subsection{Derivation of the half-width of the probability line.}

We assume that the initial state is the ground state:
\begin{multline}
\rho_0(x,y;x',y') = \psi_0(x,y)\psi_0^*(x',y')= \\ \phi_0(x)\phi_0(y)\phi_0^*(x')\phi_0^*(y'),
\end{multline}
where $\phi_0$ is the ground-state eigenfunction of a harmonic oscillator and $\psi_0$ is the initial wave function of a particle in a 2D parabolic potential.
Furthermore, we will use wave functions formalism because the purity of the initial state is $1$ and it does not change.

Wave function after $N$ non discarded measurements can be written as
\begin{equation}
\psi_N(x,y) = \frac{\cos^N(2\chi)\psi_0(x,y)}{\sqrt{P(N)}}.
\end{equation}
For this choice, the maximum of the probability density is at the $(0,0)$ point. The value of the probability density in this point is
\begin{equation}
|\psi_N(0,0)|^2 = \frac{|\psi_0(0,0)|^2}{P(N)}.
\end{equation}
We define a half width $W$ as follows:
\begin{equation}
|\psi_N(W)|^2 = \frac{|\psi_N(0,0)|^2}{2}.
\end{equation}
Here we skip $N$ in $W(N)$ and write $W(N)$ as $W$. By using the definition of $\chi$ in the main text, we write
\begin{equation}
\chi(W) = \alpha M\hbar^{-1} W/2.
\end{equation}
We assume that for a large $N$, we have $W\rightarrow0$ and  $\phi_0(0)\approx\phi_0(W)$. Thus, by using the above definitions we obtain the following equation for an unknown $W$:
\begin{equation}
\cos^{2N}(2\chi(W)) = 1/2,
\end{equation}
which can be solved by using the expansion: $\cos^{2N}(2\chi(W))\approx 1 -4 N\chi^2(W)$, valid at $W\rightarrow0$.


\begin{figure}[t]
	\begin{center}
		\includegraphics{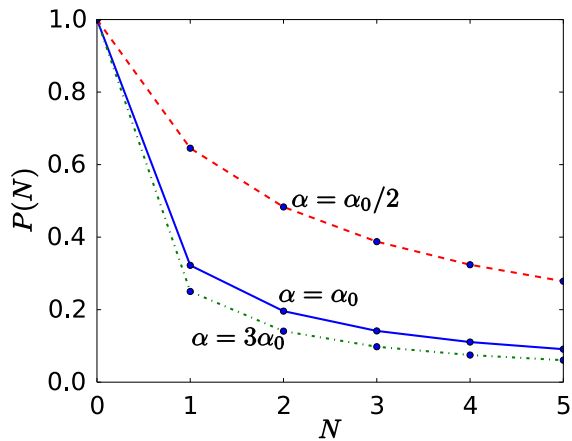}
	\end{center}
	\caption{Survival probability for compressed {\em Lotus-seed state}. Here the symbol $N$ means $N$ measurements 
	with the initial applied electric field direction and $N$~measurements with the opposite one.}
	\label{figA1}
\end{figure}
In Fig.~\ref{figA1} we present the survival probability for the ``Lotus-seed'' compressing case. As can be seen from comparison of Figs.\ref{fig2} and~\ref{figA1}, survival probability for the lotus-seed state is lower than that for the linear compression.

\subsection{``Schr\"odinger cat'' state}
If the initial state of a particle is the ground state then we can 
achieve ``Schr\"odinger cat'' state by a single-shot measurement at time $\omega\tau = \pi/4$. 
In Fig.~\ref{figA2}, we present the probability density after this single-shot spin measurement. 
The distance $L$ between two separated peaks of probability is proportional to the spin-orbit 
coupling, i.e., $L\propto\alpha$. Note that the probability density in Fig.~\ref{figA2} corresponds to a pure state of the particle.

\begin{figure}[t]
	\begin{center}
		\includegraphics{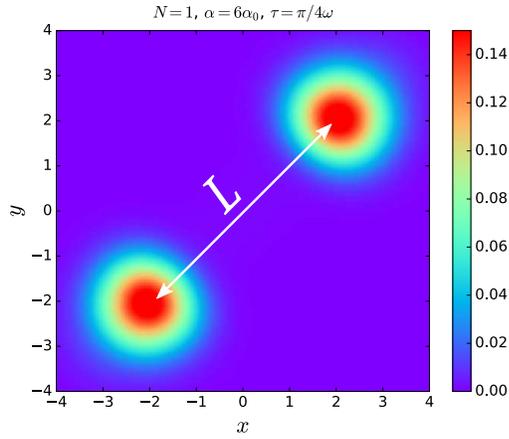}
	\end{center}
	\caption{``Schr\"odinger cat'' state corresponding to the probability density distribution
	         after one successful measurement at the special time $\tau = \pi\omega^{-1}/4$. 
	         Here spin-orbit coupling is strong: $\alpha = 6\alpha_0.$}
	\label{figA2}
\end{figure}


\begin{thebibliography}{10}
\bibitem{rev_qd}
L. P. Kouwenhoven, D. G. Austing, and S. Tarucha, Rep. Prog. Phys. \textbf{64}, 701 (2001).

\bibitem{trapped_review_1}
R. Blatt and C. F. Roos, Nature Physics \textbf{8}, 277 (2012).

\bibitem{bec_book}
L. Pitaevskii and S. Stringari, {\em Bose-Einstein Condensation}, Clarendon, Oxford (2003).

\bibitem{bec-qc}
T. Byrnes, K. Wen, and Y. Yamamoto, Phys. Rev. A \textbf{85}, 040306(R) (2012).

\bibitem{Nielsen2000}
M. A. Nielsen and I. L. Chuang, {\em Quantum Computation and Quantum
  Information} (Cambridge University Press, 2000).

\bibitem{rashba_2016}
E. I. Rashba, J. Phys.: Condens. Matter \textbf{28}, 421004 (2016)
  
\bibitem{bec_so_1}
Y. -J. Lin, K. Jimenez-Garcia and I. B. Spielman, Nature \textbf{471}, 83 (2011).

\bibitem{soc_book}
R. Winkler, {\em Spin-Orbit Coupling Effects in Two Dimensional
Electron and Hole Systems} (Springer,
Berlin, 2003).

\bibitem{note1} We use word ``purification'' in the same sence as in Ref. [\onlinecite{Hiromichi_Main_method}]: 
the physical process which transforms some quantum system from given mixed initial state to a pure state.

\bibitem{Chuang_logical_labeling}
L. Chuang, N. Gershenfeld, M. G. Kubinec and D. W. Leung, Proc. R. Soc. Lond. A \textbf{454}, 447 (1998).

\bibitem{Chuang_temporal_labeling}
E. Knill, I. L. Chuang, and R. Laflamme, Phys. Rev. A \textbf{57}, 3348 (1998).

\bibitem{ion_coherent}
J. Alonso, F. M. Leupold, B. C. Keitch and J. P. Home, New J. Phys. {\bf 15}, 023001 (2013).

\bibitem{newadd_Lo}
H.-Y. Lo,	D. Kienzler,	L. de Clercq,	M. Marinelli,	V. Negnevitsky,	B. C. Keitch, and J. P. Home, Nature \textbf{521}, 336 (2015)

\bibitem{newadd_Owen}
E. T. Owen, M. C. Dean, and C. H. W. Barnes, Phys. Rev. A \textbf{89}, 032305 (2014)

\bibitem{newadd_Douglas}
J. S. Douglas and K. Burnett, Phys. Rev. A \textbf{86}, 052120 (2012)	

\bibitem{Hiromichi_Main_method}
H. Nakazato, T. Takazawa and K. Yuasa, Phys. Rev. Lett. \textbf{90}, 060401 (2003).

\bibitem{Li2011}
Y. Li, L.-A. Wu, Y.-D. Wang, and L.-P. Yang, Phys. Rev. B {\bf 84},  094502 (2011).

\bibitem{Wu04}
L.-A. Wu, D. A. Lidar, and S. Schneider, Phys. Rev. A \textbf{70}, 032322 (2004)

\bibitem{Xu2014}
J.-S. Xu,	M.-H. Yung,	X.-Y. Xu,	S. Boixo,	Z.-W. Zhou,	C.-F. Li, A. Aspuru-Guzik and G.-C. Guo, Nat Photon {\bf 8},  113  (2014).

\bibitem{Hamiltonian_proof_1}
B. A. Bernevig, J. Orenstein, and Shou-Cheng Zhang, Phys. Rev. Lett. {\bf 97}, 236601 (2006)

\bibitem{Hamiltonian_proof_2}
J. Schliemann, J. C. Egues, and D. Loss, Phys. Rev. Lett. {\bf 90}, 146801 (2003)

\bibitem{note_spin_measurement}
Note, the process of spin measurement differs from one described in
K. C. Nowack, F. H. L. Koppens, Yu. V. Nazarov, and
L. M. K. Vandersypen, Science \textbf{318}, 1430 (2007).

\bibitem{eigen}
Another possibility is eigenstate which is superposition of number states $\sum_jc_j\ket{n_{xj}, n_{yj}}$ where $k_j = n_{xj} + n_{yj}$ 
has the same parity for all~$j$.

\bibitem{purity}
In the given example we can express purity as $\mathrm{Tr}\rho_{cN}^2 = (A_N^2 + B_N^2)/(A_N+B_N)^2$, 
where $A_N = \int\limits_{-\infty}^{\infty}\cos^{2N}(\alpha y)\varphi_0^2(y)dy$, 
$B_N = \int\limits_{-\infty}^{\infty}\cos^{2N}(\alpha y)\varphi_1^2(y)dy$, and $\varphi_{0,1}(y)$ 
are zero and first spatial wave eigenfunctions of harmonic oscillator. 
As can be seen, $B_N\propto N^{-3/2} + \exp(-(\pi/\alpha)^2)N^{-1/2} + \dots$ and 
first term of $B_N$ goes to zero faster than $A_N\propto N^{-1/2}$ while second and next 
terms of $B_N$ could be made exponentially small by choosing proper small $\alpha$, 
so in the limit $N\rightarrow\infty$ and small but fixed $\alpha\rightarrow0$ ($\alpha\neq0$) 
we have purity $\mathrm{Tr}\rho_{cN}^2\rightarrow 1$.

\bibitem{Wu_spin_measurement}
L.-A. Wu, J. Phys. A: Math. Theor, {\bf 44}, 325302 (2011); S. Oh, L.-A. Wu, Y. P. Shim, J. Fei, M. Friesen, X. Hu, Phys. Rev. A \textbf{84}, 022330 (2011).

\bibitem{Winkler}
R. Winkler, \textit{Spin-Orbit Coupling Effects in Two-Dimensional Electron and Hole Systems}, 
Springer Tracts in Modern Physics, \textbf{191} (2003).

\bibitem{Dresselhaus}
G. Dresselhaus, Phys. Rev. \textbf{100} 580 (1955).

\bibitem{Rashba}
Yu. A. Bychkov and E. I. Rashba, J. Phys. C 17, 6039 (1984).

\bibitem{bec_spin_measurement}
R. Schmied, J.-D. Bancal, B. Allard, M. Fadel, V. Scarani, P. Treutlein, N. Sangouard, Science \textbf{352}, 441 (2016).

\bibitem{SOC_in_wire}
C. Fasth, A. Fuhrer, L. Samuelson, V. Golovach, and D. Loss,
Phys. Rev. Lett. \textbf{98}, 266801 (2007)

\bibitem{Fidelity}
R. Jozsa, Journal of Modern Optics, \textbf{41}, 2315 (1994).

\bibitem{cat_state}
C. Monroe, D. M. Meekhof, B. E. King, D. J. Wineland, Science \textbf{272}, 1131 (1996).

\bibitem{Sherman}
E. Ya. Sherman and D. Sokolovski, New J. Phys. \textbf{16}, 015013, (2014).


\bibitem{about_relax_time}
C. Kloeffel and D. Loss, Annual Review of Condensed Matter Physics
{\bf 4}, 51 (2013)

\bibitem{about_relax_time_trap}
D. J. Wineland, C. Monroe, D. M. Meekhof, B. E. King, D. Leibfried, W. M. Itano, J. C. Bergquist, 
D. Berkeland, J. J. Bollinger and J. Miller, Proc. R. Soc. London Ser. A {\bf 454}, 411 (1998)


\end{thebibliography}



\end{document}